\documentclass[12pt]{article}
\usepackage{graphicx}
\usepackage{a41}
\usepackage{cite}
\usepackage{amsmath}
\usepackage{hyperref}
\usepackage{color}
\usepackage{colordvi}
\usepackage{float} %
\usepackage{lscape} %
\usepackage{amssymb} %
\usepackage{verbatim} %

\usepackage{pslatex}
\usepackage{txfonts}
\usepackage[latin1]{inputenc}
\usepackage[T1]{fontenc}

\newcommand{\N}{\nonumber}

\newcommand{\beq}{\begin{equation}}
\newcommand{\eeq}{\end{equation}}
\newcommand{\bea}{\begin{eqnarray}}
\newcommand{\eea}{\end{eqnarray}}

\newcommand{\MeV}{\rm MeV}

\newcommand{\gsim}{\raisebox{-0.07cm}{$\, \stackrel{>}{{\scriptstyle
\sim}}\, $}}

\graphicspath{/usr1/hboett/plots/qcd_pol/}
\graphicspath{./}

\begin{document}
\begin{titlepage}

\begin{flushleft}
DESY 13--202  \hfill 
\\ 
DO--TH 13/29  \\
SFB/CPP-13-87 \\
LPN-13--087   \\
November 2013 \\
\end{flushleft}
\vspace{1.5cm}
\noindent
\begin{center}
{\LARGE\bf New Exclusion Limits on Dark Gauge Forces} 

\vspace{2mm}
{\LARGE\bf from Proton Bremsstrahlung}

\vspace{2mm}
{\LARGE\bf in Beam-Dump Data }
\end{center}
\begin{center}

\vspace{2cm}

{\large Johannes Bl\"umlein$^1$ and J\"urgen Brunner$^{2}$}

\vspace{1.5cm}
{\it $^1$~Deutsches Elektronen--Synchrotron, DESY,\\
Platanenallee 6, D--15738 Zeuthen, Germany}

\vspace{4mm}
{\it $^2$~CPPM, Aix-Marseille Universit\'{e}, CNRS/IN2P3, Marseille, France}

\vspace{2.5cm}
\end{center}

\begin{abstract}
\noindent
We re-analyze published proton beam dump data taken at the U70 accelerator at IHEP
Serpukhov with the $\nu$-calorimeter I experiment in 1989 to set mass-coupling limits
for dark gauge forces. The corresponding data have been used for axion and light Higgs
particle searches in Refs.~\cite{Blumlein:1990ay,Blumlein:1991xh} before. More recently,
limits on dark gauge forces have been derived from this data set, considering a dark 
photon production from $\pi^0$-decay~\cite{Blumlein:2011mv}. Here we determine extended 
mass and coupling exclusion bounds for dark gauge bosons ranging to masses  $m_{\gamma'}$ 
of $624~\MeV$ at admixture parameters $\varepsilon \simeq 10^{-6}$ considering 
high-energy Bremsstrahlung of the $U$-boson off the initial proton beam and different 
detection mechanisms. 
\end{abstract}
\end{titlepage}

\vfill
\newpage
\sloppy
\section{Introduction}

\vspace*{4mm}
\noindent
Beyond the forces of the $SU_{3,c} \times SU_{2,L} \times U_{1,Y}$ Standard Model (SM) 
other $U_1$-fields, very weakly coupling to ordinary matter, 
may exist ~\cite{FAYET,Holdom:1985ag,Fayet:1990wx,Redondo:2010dp,Andreas:2010tp,FELDMAN,Essig:2010gu,MILLI}.
The corresponding extended Lagrangian reads \cite{Bjorken:2009mm,Essig:2010gu}
\begin{eqnarray}
\label{eq:1}
{\cal L} = {\cal L}_{\rm SM}
- \frac{1}{4} X_{\mu\nu} X^{\mu\nu}
+ \frac{\epsilon}{2} X_{\mu\nu} F^{\mu\nu} + e_\psi \epsilon \overline{\psi}
\gamma_\mu \psi X^\mu
+ \frac{m_{\gamma'}^2}{2} X_{\mu} X^{\mu}~,
\end{eqnarray}
with $X^{\mu}$ the new vector potential and $X^{\mu\nu} = \partial^\mu
X^{\nu} - \partial^\nu X^{\mu}$ the corresponding field strength tensor, and  
$F^{\mu\nu}$ the $U(1)_Y$ field strength tensor. The mixing of the new $U(1)$ and
$U(1)_Y$ of the Standard Model is induced by loops of heavy particles coupling to
both fields \cite{Holdom:1985ag,Andreas:2010tp}. We assume minimal coupling  for $X_{\mu}$ 
to all charged Standard Model fermions $\psi$, with effective charge $e_\psi \epsilon \equiv 
\hat{e}$, and $e_\psi$ being the fermionic charge under $U(1)_{\rm QED}$. For the generation 
of the mass term we assume the Stueckelberg formalism \cite{Stueckelberg:1938}, as one 
example.\footnote{Other mechanisms are possible as well, cf. e.g.\cite{Bjorken:2009mm,Blumlein:2011mv}.}

In the mass range of $m_{\gamma'} \gsim 1 \MeV$ searches for a new $U(1)$-boson have been  
performed analyzing the anomalous magnetic moments of the electron and muon \cite{Pospelov:2008xx}, 
$\Upsilon(3S)$-decays \cite{UPS}, Belle \cite{BELLE}, $J/\psi$-decays \cite{JPSI}, $K$-decays 
\cite{Beranek:2012ey}, data from KLOE-2 \cite{Babusci:2012cr},
A1 \cite{Merkel:2011ze}, APEX \cite{Abrahamyan:2011gv}, HADES \cite{Agakishiev:2013jla},
 as well as searches in electron and proton 
beam dump 
experiments, as E774 \cite{E774}, E141 \cite{E141}, E137 \cite{E137,Essig:2010xa}, Orsay 
\cite{Davier:1989wz}, KEK 
\cite{Konaka:1986cb}, $\nu$-CAL~I \cite{Blumlein:2011mv}, NOMAD and PS191 \cite{Gninenko:2011uv}, CHARM 
\cite{Gninenko:2012eq}, SINDRUM \cite{Gninenko:2013sr}, and WASA \cite{Adlarson:2013eza}. Furthermore, 
limits were derived from supernovae cooling~\cite{Turner:1987by,LSND,Batell:2009di,Bjorken:2009mm,Dent:2012mx}. 
Possibilities to search for dark photons in low energy $ep$- \cite{Freytsis:2009bh} and $e^+e^-$-scattering
\cite{KATZIN} have been explored. Effect of massive photons on the $\mu$-content of air showers were studied 
in \cite{Ebr:2013iea}. Updated summaries of exclusion limits and reactions have been given 
in 
Refs.~\cite{Andreas:2012xh,Andreas:2012mt,Reece:2009un,FRASCATI,Gninenko:2013rka,Gershtein:2013iqa}. The present limits 
in the 
$m_{\gamma'}-\epsilon$ plane range from $\epsilon \in [5 \times 10^{-9}, 10^{-2}]$ and a series of mass regions 
in $m_{\gamma'} \in [2~m_e, \sim 3$~GeV], with an unexplored range towards lower values of $\epsilon$ and larger 
masses.

In the present note we derive new exclusion bounds on dark $\gamma'$-bosons using
proton beam dump data at $p \sim 70$~GeV, based on potential $\gamma'$-Bremsstrahlung off the 
incoming proton beam searching for electromagnetic showers in a neutrino calorimeter \cite{Barabash:2002zd}.
In a previous analysis \cite{Blumlein:2011mv} exclusion limits were derived based on $\gamma'$-production in 
the decay of the $\pi^0$-mesons. These beam-dump data have been used in the in axion \cite{AXION} and light 
Higgs boson searches, cf.~\cite{PROP,Blumlein:1990ay,Blumlein:1991xh}, in the past.

In the following we first derive the production cross section, describe the detection process, the 
experimental set-up and data taking, and then derive new exclusion limits on the mass and coupling of a 
hypothetic $U_1'$-boson.
\section{Production Cross Section}

\vspace*{1mm}
\noindent
One production channel for a $U_1'$-boson $\gamma'$ in a high-energy proton 
beam dump is given by small-angle initial-state radiation from the incoming proton 
at large longitudinal momentum, followed by a hard proton-nucleus interaction.
The hadronic cross section is used in form of a parameterization of the measured 
distributions. Corresponding radiator functions may be derived using the 
Fermi-Williams-Weizs\"acker method~\cite{Fermi:1924tc,Williams,vonWeizsacker:1934sx} to good 
approximation\footnote{For a review see \cite{Kessler:1975} and references therein. Early 
applications are found in \cite{Kessler:1960,LU}.}. For the derivation often old-fashioned perturbation 
theory~\cite{Weinberg:1966jm} in the infinite momentum frame is used in the literature, 
cf.~\cite{Baier:1973ms,Chen:1975sh,Altarelli:1977zs}. As well known, the corresponding radiators, 
beyond the universal contributions being free of mass effects, are {\it no} generalized splitting 
functions and are {\it not} process independent\footnote{Cf., however, Ref.~\cite{Martin:1996eva}.}. 
They just describe a factorizing weight-function of a differential cross section $d \sigma_a$ relative 
to a sub-process given by $d \sigma_b$, 
\begin{eqnarray}
d \sigma_a = w_{ba}(z,p_\perp^2) dz dp_\perp^2~d \sigma_b,
\end{eqnarray}
cf.~Refs.~\cite{Chen:1975sh,Altarelli:1977zs}. 

The Fermi-Williams-Weizs\"acker approximation was also derived using covariant 
methods, cf.~\cite{Kessler:1975} and \cite{Kim:1973he,Frixione:1993yw}. Here 
one may consider the splitting-vertex $p \rightarrow \gamma' + p'$ only 
\cite{Baier:1973ms,Chen:1975sh,Altarelli:1977zs,Frixione:1993yw}, which will 
lead to finite fermion mass corrections up to $\sim M^2$ in the fermion mass. 
Using the
method of \cite{Altarelli:1977zs} and accounting for a finite fermion mass one 
reproduces the results given in 
\cite{Baier:1973ms,Chen:1975sh,Frixione:1993yw}\footnote{In case of the 
representation given in \cite{Frixione:1993yw} the denominators containing 
$p_\perp^2$ are obtained from the virtuality $q^2$ in the deep-inelastic case 
for small angles $\theta^2 \ll 1$, where $
q^2 = - \left[M^2 z^2 + (A^2 -M^2)/(4A^2) \theta^2\right]/(1-z) 
\equiv - (z^2 M^2 + p_\perp^2)/(1-z)$~, with $A = E(1+\beta)(1-z)$, $z \equiv 
y_{\rm BJ}$, $\beta = (1-M^2/E^2)^{1/2}$, and $E$ the energy of the 
incoming fermion beam.}. 

A more general approach, the generalized Fermi-Williams-Weizs\"aecker method, 
relies on the scattering process 
\begin{eqnarray}
\label{eq:REAC}
b + p \rightarrow \gamma' + p',
\end{eqnarray}
with $b$ the boson being exchanged between in the incoming fermion and 
the hadronic target, for which we assume $b$ being a vector, cf. also \cite{Frixione:1993yw}. 
Following~\cite{Kim:1973he} 
the contraction of the fermionic tensor corresponding to (\ref{eq:REAC}) with the incoming 
target momentum $P_{i,\mu}$ is given by
\begin{eqnarray}
\frac{L^{\mu\nu} P_{i,\mu} P_{i,\nu}}{M_i^2} 
= 
  \frac{q_z^2}{(q_z - q_0)^2} \left(L_{00} + L_{zz} - 2 L_{0z}\right)
+ \frac{q_\perp^2}{(q_z - q_0)^2} \left(
  \cos^2\varphi L_{xx} 
+ \sin^2\varphi L_{yy} \right),
\label{eq:L1}
\end{eqnarray}
where $M_i$ denotes the target mass and $q_z, q_\perp$ are the components
of the momentum of the boson $b$. As shown in \cite{Kim:1973he} the terms 
$L_{00} + L_{zz} - 2 L_{0z}$ are strongly suppressed relative to those of the 
second term. The dominant contribution to (\ref{eq:L1}) 
stems from the region of very small values of $q_\perp^2$ and one may rewrite 
this relation performing the $\varphi$-integral as
\begin{eqnarray}
\frac{1}{2\pi} \int_0^{2\pi} d\varphi 
\frac{L^{\mu\nu} P_{i,\mu} P_{i,\nu}}{M_i^2} 
\approx \frac{q_\perp^2}{(q_z - q_0)^2} \left(-\frac{1}{2} g_{\mu\nu} L^{\mu\nu} 
\right)_{q^2 = q^2_{\rm min}},
\label{eq:L2}
\end{eqnarray}
since $L_{00} \approx L_{zz}$. In the following the virtuality $q^2$ is set 
effectively to zero. 

We consider $b$ as a vector particle and $\gamma'$ as the $U_1'$-gauge boson
with mass $m_{\gamma'}$. 
The matrix element $\overline{|{\cal M}|^2}$ averaging over the initial state
spins is given by
\begin{eqnarray}
\overline{|{\cal M}|^2} = -\frac{1}{8} g^{\mu\nu} L_{\mu\nu} 
                        &=&     - \frac{S}{U} - \frac{U}{S}
                                 +  2(2M^2 +m_{\gamma'}^2) 
                                 \left(\frac{1}{S} + \frac{1}{U}\right)
                                 + 4 M^4 \left(\frac{1}{S} + \frac{1}{U}\right)^2
\N\\  &&                        
                                 +  2 M^2 m_{\gamma'}^2 
                                 \left(\frac{1}{S^2} + \frac{1}{U^2}\right)
                         - 2 \frac{m_{\gamma'}^4}{S~U}~,
\label{eq:MAT}
\end{eqnarray}
with the projector $-g_{\mu\nu} + k_\mu k_\nu/m_{\gamma'}^2$ for the polarization sum
for the $U_1'$-boson, is easily calculated using {\tt FORM} \cite{Vermaseren:2000nd}.
Since we now refer to the the $2 \rightarrow 2$ scattering process (\ref{eq:REAC}) 
also fermion mass terms up to $\sim M^4$ contribute. Here we have not specified 
the nature \cite{Proca:1936,Stueckelberg:1938} of the produced boson. Due to 
the production of a massive final state boson $\gamma'$ three degrees of 
polarization contribute. This, however, does not lead to $1/m_{\gamma'}^k$-terms, with $k > 
0, k~\in~\mathbb{N}$, in (\ref{eq:MAT})\footnote{As has been discussed in the literature extensively
\cite{CJ:1950,Coester:1951,JR:1976,Ruegg:2003ps} the transition in scattering 
cross sections from a massive boson to the massless limit needs not always 
to be continuous.}. Massive boson production in Bremsstrahlung has also been 
considered e.g. in \cite{Linsker:1972dn,Akhundov:1976yy} and for massless 
fermions in \cite{Bjorken:2009mm}.  

The invariants $S$ and $U$ in (\ref{eq:MAT}) are given by
\begin{eqnarray}
U &=& u - M^2 = (p-k)^2 - M^2  = m_{\gamma'}^2 - 2 p.k~,  \\
S &=& s - M^2 = (p'+k)^2 - M^2 = m_{\gamma'}^2 + 2 p'\!\!.k~,
\end{eqnarray}
with $p,p'$ and $k$ the momenta of the incoming, outgoing fermion and produced
boson $\gamma'$. From the matrix element in Eq.~(\ref{eq:MAT}) we derive the 
splitting probability for the process $P \rightarrow \gamma' + P'$ and set
the momentum of the boson $b$ to $q = 0$. Referring to the infinite momentum frame given 
by the fast moving incoming fermion of momentum  $P$ the 4--momenta are given by 
\cite{Chen:1975sh,Altarelli:1977zs}
\begin{eqnarray}
p &=& \left(P + \frac{M^2}{2P}; P, 0, 0\right)\\
k &=& \left(zP + \frac{p_\perp^2 + m_{\gamma'}^2}{2Pz}; zP, p_x, p_y\right)\\
p'&=& \left((1-z)P + \frac{M^2 + p_\perp^2}{2P(1-z)}; (1-z)P, -p_x,- p_y\right)~.
\end{eqnarray}
The invariants read, cf. also \cite{Bjorken:2009mm},
\begin{eqnarray}
U =  - \frac{1}{z} \left[ (1-z) m_{\gamma'}^2  + z^2 M^2 + p_\perp^2\right],~~~~ 
S &=&  - \frac{U}{1-z}~.
\end{eqnarray}
One thus obtains
\begin{eqnarray}
\label{eq:MAT1}
w_{ba}(z, p_\perp^2) dz dp_\perp^2 &=& \frac{\alpha'}{2\pi} \Biggl\{
\frac{1 + (1-z)^2}{z} - 2z(1-z) \left[\frac{2 M^2 + m_{\gamma'}^2}{H} - z^2  \frac{2 
M^4}{H^2}  \right] \N\\
&& + 2 z (1-z) [1 + (1-z)^2] \frac{M^2 m_{\gamma'}^2}{H^2} + 2 z (1-z)^2 
\frac{m_{\gamma'}^4}{H^2} \Biggr\} \frac{dz dp_\perp^2}{H}~,
\end{eqnarray}
with  $\alpha' = (\hat{e})^2/(4 \pi)$ and
\begin{eqnarray}
\label{eq:MAT1A}
H(p_\perp,z)  = p_\perp^2 + (1-z) m_{\gamma'}^2  + z^2 M^2~.
\end{eqnarray}
The first term in (\ref{eq:MAT1}) denotes the well-known splitting function
$P_{\gamma' f}(z)$. In the limit $M^2 \rightarrow 0$~~Eq.~(\ref{eq:MAT1}) agrees with 
a corresponding expression in \cite{Bjorken:2009mm}.

The $p_\perp^2$-integral in (\ref{eq:MAT1}) is regularized by both masses 
$m_{\gamma'}$ and $M$ individually. It is given by
\begin{eqnarray}
\label{eq:MAT1pt}
w_{ba}(z) dz 
&=& 
\frac{\alpha'}{2\pi} \Biggl\{
\frac{1 + (1-z)^2}{z} \ln\left(1 + \frac{p_{\perp,{\rm max}}^2}{A}\right) 
- 2z(1-z) (2 M^2 + m_{\gamma'}^2) \frac{p_{\perp,{\rm max}}^2}{A(p_{\perp,{\rm max}}^2 + A)} 
\N\\
&&
+  2 z (1-z) \left[2 z^2 M^4 
       + [1 + (1-z)^2] M^2 m_{\gamma'}^2 
       + (1-z) m_{\gamma'}^4 \right] \frac{p_{\perp,{\rm max}}^2(p_{\perp,{\rm max}}^2+2A)}
{2(p_{\perp,{\rm max}}^2 + A)^2 A^2}
\Biggr\} dz~,
\end{eqnarray}
with $A = (1-z) m_{\gamma'}^2 + z^2 M^2$.

The final production cross section reads
\begin{eqnarray}
\label{eq:MAT3}
\sigma_{p+A \rightarrow \gamma' + X} &=& \int_{z_{\rm min}}^{z_{\rm max}} dz 
\int_0^{p_{\perp, \rm max}^2} dp_\perp^2~{w}_{\gamma' p}(z, p_\perp^2)  
\sigma_{pA}(s') \theta[f(z,p_\perp^2)]~,
\end{eqnarray}
with $s' = (M+E_p)^2 (1-z)$, $E_p$ the beam energy of the accelerator,
$\sigma_{pA}(s')$ the hadronic scattering 
cross section after  $U_1'$-boson emission and $\theta[f(z,p_\perp^2)]$ summarizing the 
experimental cut conditions. The cross section $\sigma_{pA}(s')$ is related to the $pN$-scattering cross 
section by a function $f(A)$, which drops out again in calculating the event rate.
The inelastic scattering cross section $\sigma_{pp}$ is taken from experimental 
data,~cf.~Ref.~\cite{Nakamura:2010zzi}~:
\begin{equation}
\sigma_{pp}(s') = Z + B\cdot \log^2\left(\frac{s'}{s_0}\right) + Y_1\left(\frac{s_1}{s'}\right)^{\eta_1} - 
Y_2\left(\frac{s_1}{s'}\right)^{\eta_2},
\end{equation}
where $Z = 35.45$~mb, $B = 0.308$~mb, $Y_1 = 42.53$~mb, $Y_2 = 33.34$~mb, $\sqrt{s_0} = 5.38$~GeV, 
$\sqrt{s_1} = 1$~GeV, $\eta_1 = 0.458$ and $\eta_2 = 0.545$.
\restylefloat{figure}
\begin{center}
\begin{figure}[H] 
\begin{center}
\includegraphics[width=0.8\linewidth]{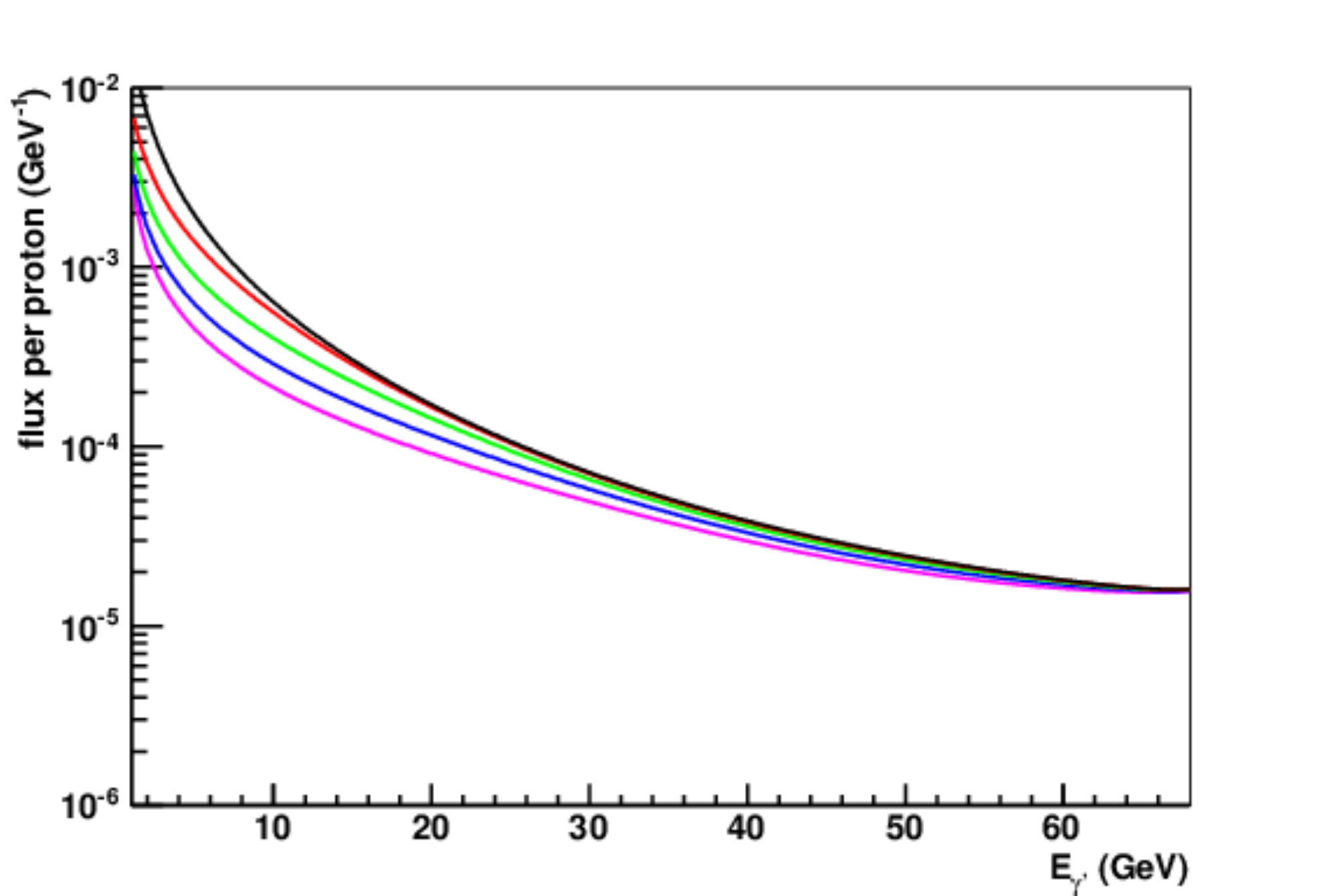}
\end{center}
\caption[]{
\small
Flux of produced $\gamma'$-particles within the angular acceptance of the detector per beam proton
as function of their energy in the laboratory frame. The black, red, green, blue and magenta lines 
correspond to
$\gamma'$ with masses between 0 and 800~MeV in steps of 200~MeV for $\epsilon = 1$.}
\label{FIG:1}
\end{figure}  
\end{center}

Finally we would like to briefly summarize the condition of use for the 
Fermi-Williams-Weizs\"acker approximation given in 
\cite{Kim:1973he,Kim:1972gw} for the present set-up. These are
\begin{eqnarray}
E^2          &\gg& (p+k)^2,~~M^2 \\
E_\gamma     &\gg& M_\gamma' \\
E - E_\gamma &\gg& \Delta, M, \sqrt{p'^2}, \frac{1}{M}\left[M^2 - p'^2\right]~,
\end{eqnarray}
with $\Delta = (M_f^2 - M_i^2)/(2 M_i)$ and $M_i \equiv M$. In case of a quasi-elastic emission
of the $U_1'$-boson one expects the hadronic mass $M_f = \sqrt{p'^2}$ of similar size than the nucleon mass 
$M$. The conditions translate into
\begin{eqnarray}
\label{eq:cond1}
P^2 &\gg& z M^2 + \frac{1}{z}\left[p_\perp^2 + (1+z) m_{\gamma'}^2\right]\\
\label{eq:cond2}
P^2 &\gg& M^2 
\end{eqnarray}
\begin{eqnarray}
\label{eq:cond3}
zP  + \frac{p_\perp^2 + m_{\gamma'}^2}{2zP} &\gg& m_{\gamma'} \\
\label{eq:cond4}
(1-z) P + \frac{M^2 + p_\perp^2}{2P(1-z)} &\gg& \Delta;~M;~\sqrt{p'^2};~\frac{1}{M}\left[M^2 - p'^2\right]~.
\end{eqnarray}

Again, for quasi-elastic splitting one has $\sqrt{p'^2} \sim M$. While (\ref{eq:cond2}) is fulfilled
automatically at high energy accelerators, (\ref{eq:cond1}, \ref{eq:cond3}, \ref{eq:cond4}) set 
constraints
on $z$ in dependence of the values of $p_\perp^2$ and $m_{\gamma'}$ and have to be tested 
accordingly. 
These conditions may be summarized by 
\begin{eqnarray}
E_p, E_{\gamma'}, E_p - E_{\gamma'} \gg M, m_{\gamma'}, \sqrt{p_\perp^2}~.
\label{Eq:approx}
\end{eqnarray}
From the experimental setup one obtains $E_p = 70$~GeV and $p_{\perp}^2 < 1$~GeV$^2$ (see below).
Further we only test masses $m_{\gamma'} < 1$~GeV and we restrict to the energy
range 10~GeV~$< E_{\gamma'} <$~60 GeV, which corresponds to the condition 0.14~$< z <$~0.86.
\restylefloat{figure}
\begin{center}
\begin{figure}[H] 
\begin{center}
\includegraphics[width=0.8\linewidth]{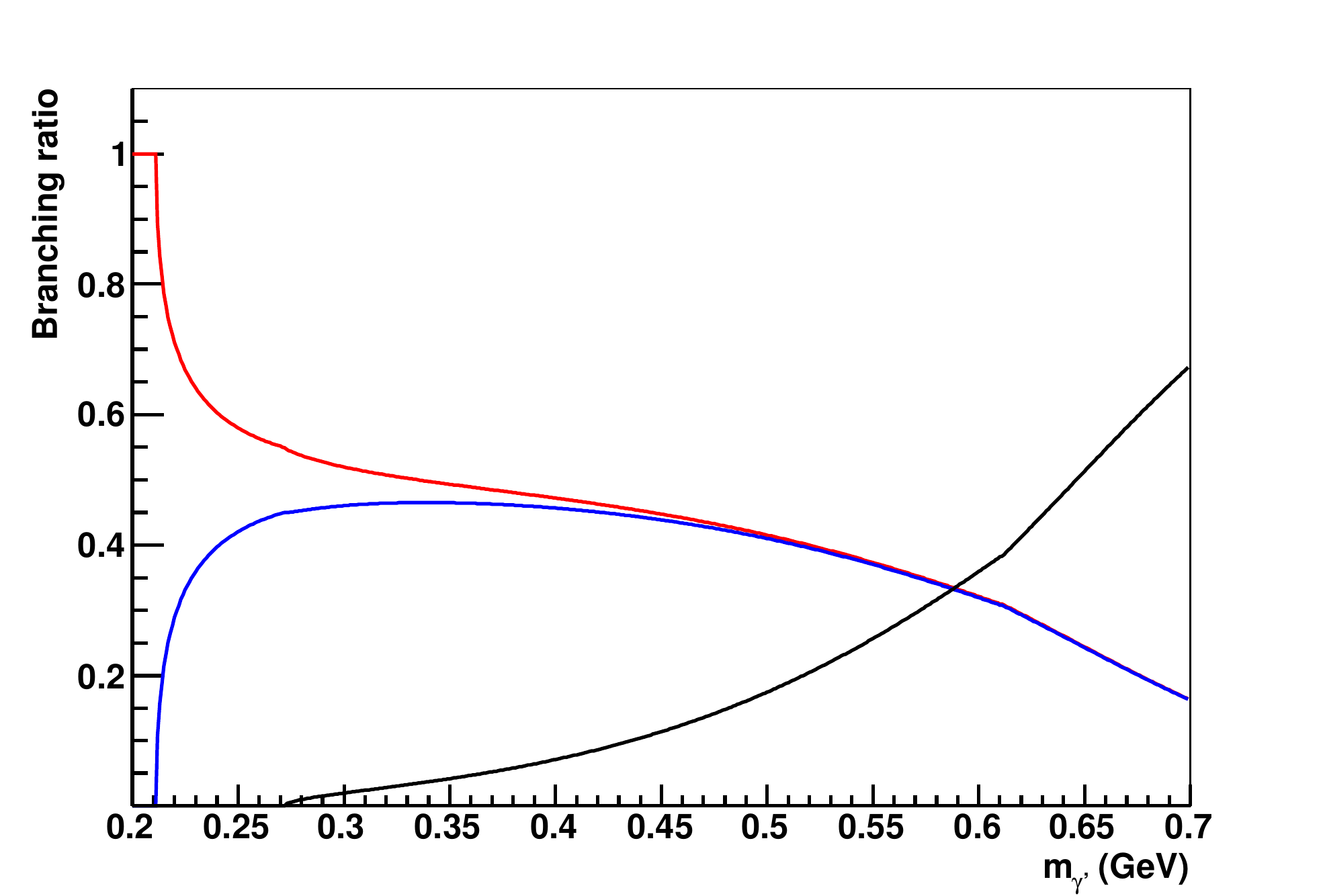}
\end{center}
\caption[]{
\small
Branching ratio of $\gamma'$ into $e^+e^-$ (red), $\mu^+\mu^-$ (blue) and hadrons (black).}
\label{FIG:2}
\end{figure}
\end{center}
This combination of constraints ensures the validity
of the approximations used according to the conditions of Eq.~(\ref{Eq:approx}).

The event rates in the detector are calculated using the differential $\gamma'$-rate per 
proton interaction
\begin{eqnarray}
\label{eq:MAT5}
\frac{dN}{dE_{\gamma'}} &=& \frac{1}{E_p}\frac{\sigma_{pA}(s')}{\sigma_{pA}(s)}
\int_0^{p_{\perp, \rm max}^2} ~{w}_{ba}(z,p_{\perp}^2) dp_{\perp}^2 ,
\end{eqnarray}
where $s' = 2M(E_p-E_{\gamma'}$) is the reduced center-of-mass 
energy after the emission of the $\gamma'$ and $s=2ME_p$. The resulting $\gamma'$-rate is 
shown in Figure~1 for five values of $m_{\gamma'}$ between $0$ and $800$~MeV and $\epsilon = 1$.
\section{The Detection Processes}

\vspace*{1mm}

In Ref.~\cite{Blumlein:2011mv} we restricted the analysis to the mass range $2m_e < m_{\gamma'} < m_\pi^0$. 
Here the only relevant decay channel is
$\gamma' \rightarrow e^+ e^-$. However, the Bremsstrahlung process can produce particles with $m_{\gamma'} > 
m_\pi^0$. Therefore we 
consider here as well the decay channels $\gamma' \rightarrow \mu^+ \mu^-$ and $\gamma' 
\rightarrow~\text{hadrons}$.
The partial decay width of the $\gamma'$-boson into a lepton pair is given by 
\cite{Essig:2010gu}
\begin{eqnarray}
\Gamma(\gamma' \rightarrow l^+ l^-) = \frac{1}{3} \alpha_{\rm QED} m_{\gamma'} \epsilon^2 \sqrt{1 - 
\frac{4 m_l^2}{m_{\gamma'}^2}}\left(1 + 
\frac{2 m_l^2}{m_{\gamma'}^2}\right)~,
\label{Eq:decay}
\end{eqnarray}
where $l$ indicates either a muon or an electron. 
The partial decay width into hadrons is determined following the approach having been proposed 
in~\cite{Bjorken:2009mm}
\begin{equation}
\Gamma(\gamma' \rightarrow~\text{hadrons}) = \frac{1}{3} \alpha_{\rm QED} m_{\gamma'} \epsilon^2 
\frac{\sigma(e^+e^- \rightarrow hadrons)}{\sigma(e^+e^- \rightarrow \mu^+\mu^-)}~,
\end{equation}
where the ratio of the hadron production cross section with respect to muons is taken from~\cite{Nakamura:2010zzi}.
The resulting branching ratios for the three channels are shown in Figure~\ref{FIG:2}.
For $m_{\gamma'} < 2m_\mu$ only the decay into $e^+e^-$ is allowed. For $2m_\mu < m_{\gamma'} < $~400~MeV the suppression of the muon channel compared 
to the electron
channel due to the kinematic factor in Eq.~(\ref{Eq:decay}) is visible. 
For $m_{\gamma'} >$~600~MeV the hadronic decay starts to dominate.

The $\gamma'$ decay probability $w_{\rm dec}$ inside the fiducial volume of the detector
for a leptonic decay $\gamma' \rightarrow l^+ l^-$ is given by
\begin{eqnarray}
w_{\rm dec} = 
Br(\gamma' \rightarrow l^+ l^-)
\exp\left[- \frac{l_{\rm dump} }{c \tau(\gamma')} 
\frac{m_{\gamma'}}{|\vec{k}|}\right] 
\left[1 - \exp\left(-\frac{l_{\rm fid} }{c \tau(\gamma')}\frac{m_{\gamma'}}{|\vec{k}|}
\right)\right]
~,
\end{eqnarray}
with $\tau(\gamma')$ the lifetime of the $\gamma'$ for a given mass (i.e. the inverse of the total decay 
width), 
$c$ the velocity of light, $m_{\gamma'}$ and $\vec{k}$ are the mass and 3-momentum of the $\gamma'$-boson.
$l_{\rm dump}$ denotes the distance of the fiducial volume from the beam dump and $l_{\rm fid}$ the 
length of the fiducial volume itself. 

For $m_{\gamma'} < 2m_e$ all decay channels which are discussed above are kinematically forbidden. 
The $\gamma'$-particles which traverse the detector could be detected via Bethe-Heitler 
electron-positron pair production. For the considered mass range ($m_{\gamma'} < 2m_e$) and the energy range 
10~GeV~$< E_{\gamma'} <$~60~GeV the total cross section of this process is largely independent both from 
$E_{\gamma'}$ and $m_{\gamma'}$ and can 
be related to the well known pair production process of photons. 
For the interaction length of the pair production process of a $\gamma'$ one trivially finds
\begin{equation}
\lambda^{M}_{\gamma'} = \epsilon^2 \frac{9}{7} \frac{X_0^M}{\rho^M}~,
\end{equation}
with $X_0^M$ and $\rho^M$ the radiation length and density of material $M$ respectively. 

To calculate the pair production probability inside the fiducial volume one further needs to know
the total depth $v^M$ of each material $M$ in the veto region before the detector as well as the corresponding
length $f^M$ in the fiducial area of the detector. The interaction probability can then be calculated as
\begin{eqnarray}
w_{\rm int} = 
\exp\left[- \sum_M\frac{v^M }{\lambda^{M}_{\gamma'}}\right]  
\left[1 - \exp\left(-\sum_M\frac{f^M}{\lambda^{M}_{\gamma'}}
\right)\right]2\sin^2\left(\frac{l_{dump}m^2_{\gamma'}}{4E_{\gamma'}}\right)~.
\label{Eq:pair}
\end{eqnarray}
The last term accounts for the coherent mixing of the $\gamma'$-boson with the photon in analogy to 
neutrino
oscillations~\cite{Jaeckel:2010xx}. For the present setup the effect becomes important for 
$m_{\gamma'} < 100$~eV 
but it can safely be neglected  for larger values of $m_{\gamma'}$ as the coherence length becomes too small 
and the term averages to one.

\section{The Experimental Setup and Data Taking}

\vspace*{1mm}
\noindent
The beam dump experiment was carried out at the U70 accelerator at IHEP Serpukhov
during a three months exposure in 1989. Data have been taken
with the $\nu$-CAL I experiment, a neutrino detector. All technical 
details of this experiment have been described in \cite{Blumlein:1990ay}
and a detailed description of the detector was given in \cite{Barabash:2002zd}.
Here we only summarize the key numbers which are crucial for the present analysis.
 
The target part of the detector is used as a fiducial volume to detect the decays 
of the $\gamma'$-boson. It has a modular structure and consists of 36 identical modules
along the beam direction. Each of the modules is composed of a 5~cm thick aluminum 
plate, a pair of drift chambers to allow for three dimensional tracking 
and a 20~cm thick liquid scintillator
plane to measure the energy deposit of charged 
particles.

For the beam dump experiment a fiducial volume of 30 modules with a total length of $l_{\rm fid}=23$~m 
is chosen, starting with the fourth module at a distance of 
$l_{\rm dump}=64$~m down-stream of the beam dump. Three modules in front of the fiducial volume are used as a 
veto in addition to a passive 54~m long iron shielding. This leads to the following set of 
material parameters needed for the
pair production calculation:
\begin{table}[H]
\begin{center}
\begin{tabular}{||c|c|c|c|c||}\hline
M & $\rho^M$ (g/cm$^3$)& $X_0^M$ (g/cm$^2$) & $v^M$ (m) & $f^M$ (m) \\ \hline
Aluminum & 2.699 & 24.01 & 0.15 & 1.50 \\
Liquid Scintillator & 0.703 & 45.00 & 0.60 & 6.00 \\
Iron & 7.874 & 13.84 & 54.00 & -- \\ \hline
\end{tabular}
\caption[]{Material parameters of the most important detector components.}
\end{center}
\label{TAB:1}
\end{table}


The lateral extension of the 
fiducial volume
is $2.6 \times 2.6$~$\mbox{m}^2$. In the following we use conservatively a slightly smaller
fiducial volume, defined as a cone pointing to the beam dump with a ground circle of 
2.6~m in diameter at the end of the fiducial volume, i.e. at a distance of 87~m 
from the dump.
This leads to the following simple fiducial volume cut
\begin{equation}
(p_\perp/p_L)_{\rm lab} < 1.3/87 = 0.015~.
\end{equation}
\restylefloat{figure}
\begin{center}
\begin{figure}[H] 
\begin{center}
\includegraphics[width=0.45\linewidth]{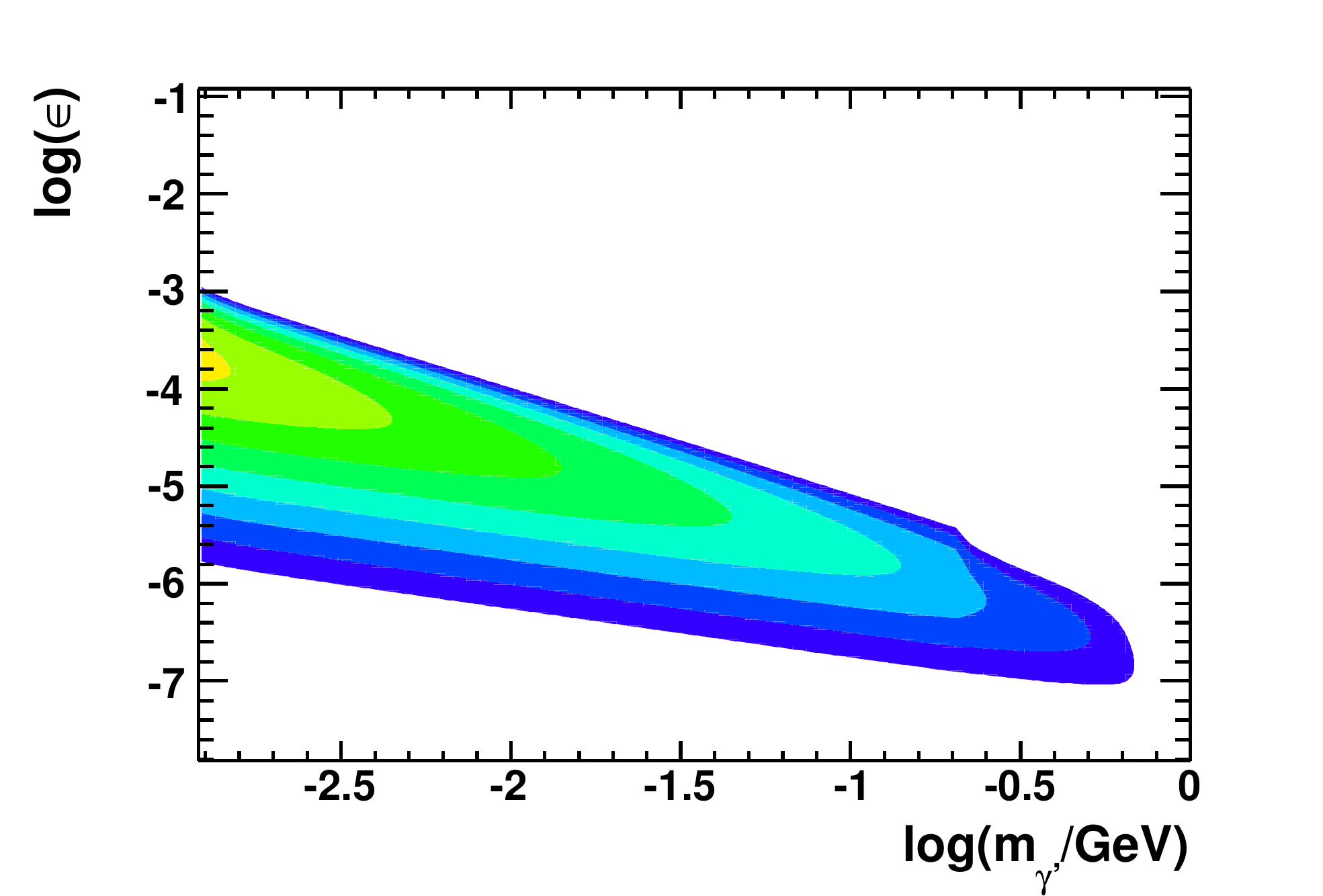}
\includegraphics[width=0.45\linewidth]{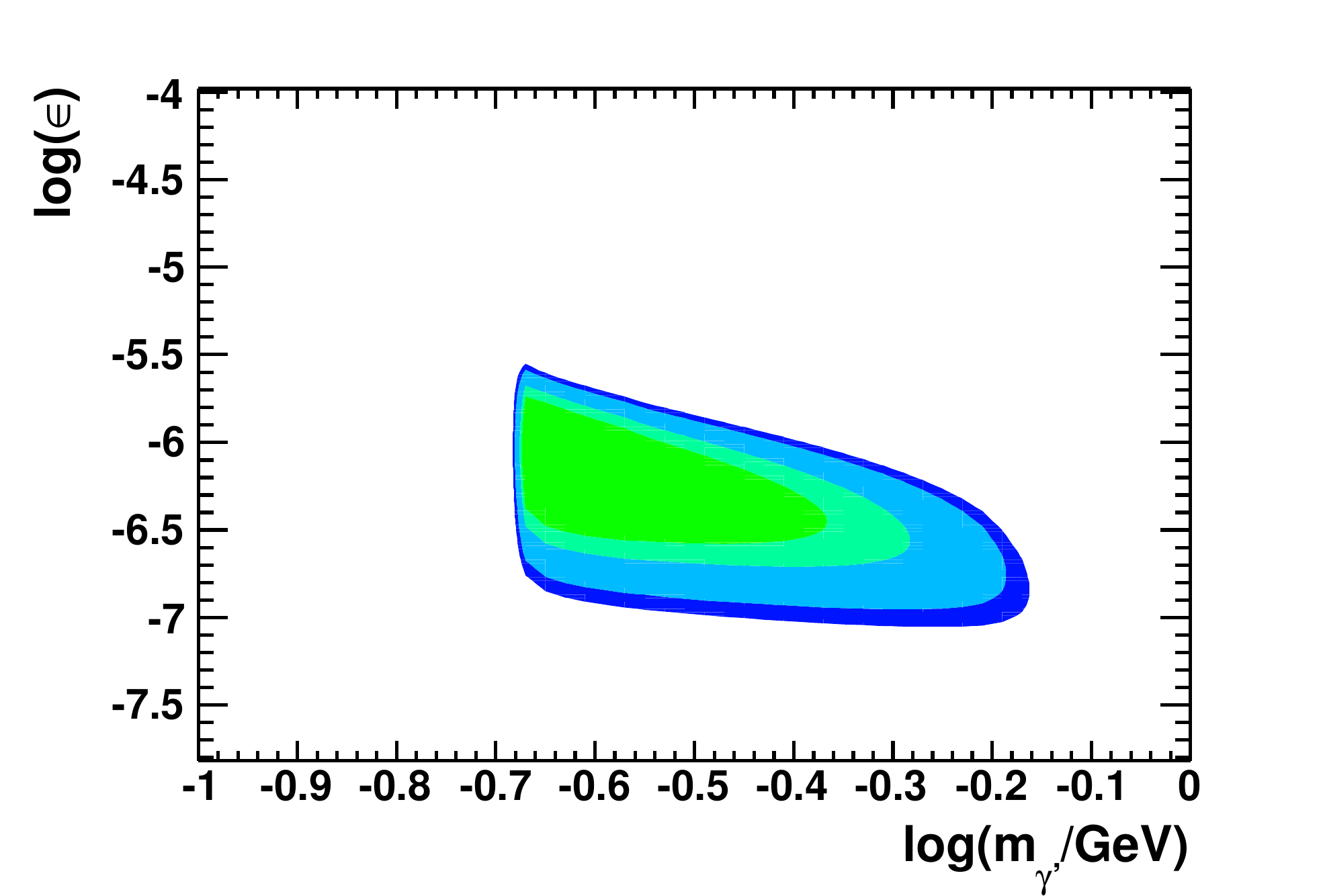}
\end{center}
\caption[]{
\small
Expected $\gamma'$-events in the electron channel (left) and muon channel (right). 
Color bands (left) per decade from $10^7$ events (yellow) to one event (dark blue) 
and (right) per semi-decade from 30 events (green) to one event (dark blue).}
\label{FIG:3}
\end{figure}
\end{center}

During the three months exposure time in 1989 $N_{\rm tot}=1.71\times 10^{18}$ protons on target had been 
accumulated~\cite{Blumlein:1990ay}. The signature of event candidates from 
$\gamma'\rightarrow e^+e^-$ is a single 
electromagnetic shower in beam direction. This signature is identical to the one from 
the axion or light Higgs particle decay search which was performed 
in~\cite{Blumlein:1990ay}. Electromagnetic showers with energies larger than 10~GeV are detected with an 
efficiency $\varepsilon_e=70\%$~\cite{Blumlein:1990ay}.
From the total data sample of 3880 reconstructed events, 1 isolated shower with $ E > $10~GeV is selected, 
which is
compatible with a background 
estimate of 0.3~events from the simulation of $\nu_\mu$ and $\nu_e$ interactions in the detector. 

In~\cite{Blumlein:1991xh} the same data set is searched for a decay signature of light Higgs or axions into 
$\mu^+\mu^-$. Again this signature is identical to the corresponding decay of a $\gamma'$ into a muon pair.
For $E_{\mu_1}+E_{\mu_2}>$10~GeV the detection efficiency is found to be $\varepsilon_\mu 
= 80\%$~\cite{Blumlein:1991xh}.
From the total data sample, one muon pair with $E_{\mu_1}+E_{\mu_2} > $10~GeV is selected, which is
compatible with a background 
estimate of 0.7~events. 

\section{Results}

\vspace*{1mm}
\noindent
The total number of expected signal events can be calculated as
\begin{equation}
N_{\rm sig} = N_{\rm tot}\times\varepsilon_l \int dE \frac{dN}{dE} w_{\rm x}(E)~.
\end{equation}
The integration is carried out over the energies of the $\gamma'$ in the range 10--60~GeV.
$w_{\rm x}$ corresponds to $w_{\rm dec}$ or $w_{\rm int}$ depending on the channel in question.
The dependence of $N_{\rm sig}$ on $m_{\gamma'}$ and $\epsilon$ for the two decay channels is shown in 
Figure~\ref{FIG:3}.

The overall shape of the contour is similar to the one obtained in~\cite{Blumlein:2011mv}. 
The maximal event numbers are about two orders of magnitude below the values found in~\cite{Blumlein:2011mv}
and the contour is narrower, both due to the lower flux from Bremsstrahlung with respect to production from 
$\pi^0$-decays. However the present contour is not limited to $m_{\gamma'} < m_{\pi^0}$ and indeed 
events are expected for masses as high as $\sim 600$~MeV. The muon channel contributes with maximally few 
tens events 
at $m_{\gamma'}=250$~MeV and $\epsilon=3\cdot 10^{-6}$. For $m_{\gamma'}>2m_\mu$ both electron and muon channels
contribute about equally, therefore the combination of these two channels will improve the sensitivity in this 
mass range.
\restylefloat{figure}
\begin{center}
\begin{figure}[H] 
\begin{center}
\includegraphics[width=0.8\linewidth]{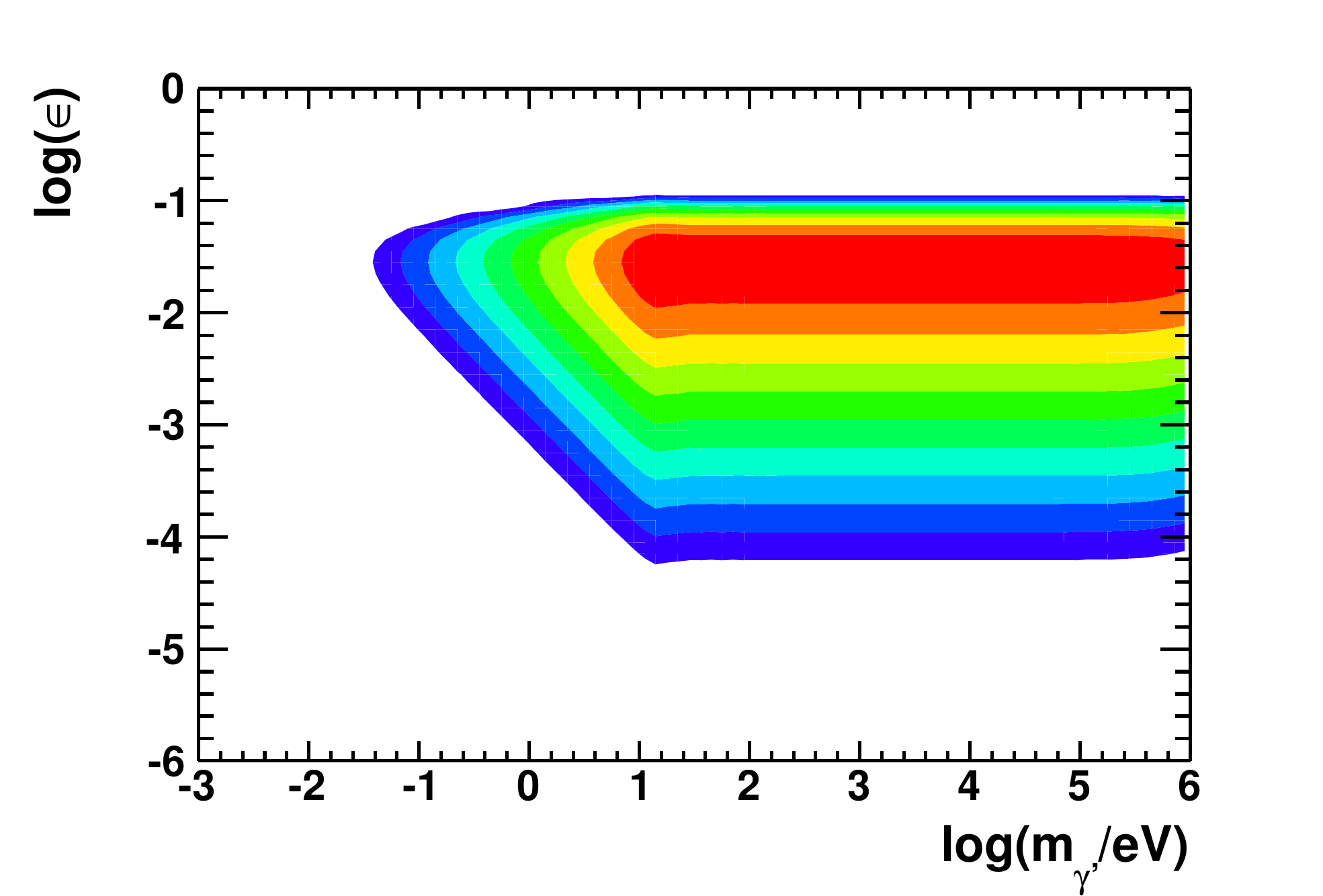}
\end{center}
\caption[]{
\small
Expected $\gamma'$ events from pair production. 
Color scale in $\log_{10}$ from $10^9$ events (red) to one event (dark blue).}
\label{FIG:4}
\end{figure}
\end{center}
Figure~\ref{FIG:4} shows the expected event numbers due the pair production process. More than $10^9$ events would be expected
over a large mass range for $\epsilon \approx 0.03$. For $\epsilon > 0.1$ the sensitivity quickly 
drops as the dark photons start to be
absorbed in the iron absorber in front of the detector as normal photons do. For $\epsilon < 10^{-4}$ the combined production and
interaction probability, which scales as $\epsilon^4$ in this range, becomes to small to produce any detectable signal. 
The narrowing of the contour for 0.01~eV$ < m_{\gamma'} <$10~eV is due to the oscillation term 
of Eq.~(\ref{Eq:pair}). 
\restylefloat{figure}
\stepcounter{footnote}           
\footnotetext{We thank S. Andreas for designing this graph. The exclusion curves for the electron-beam dump
have been recalculated in \cite{Andreas:2012xh} and are shown here. Note a difference to 
\cite{Bjorken:2009mm} in case of E137.}
\begin{center}
\begin{figure}[H] 
\begin{center}
\includegraphics[width=0.8\linewidth]{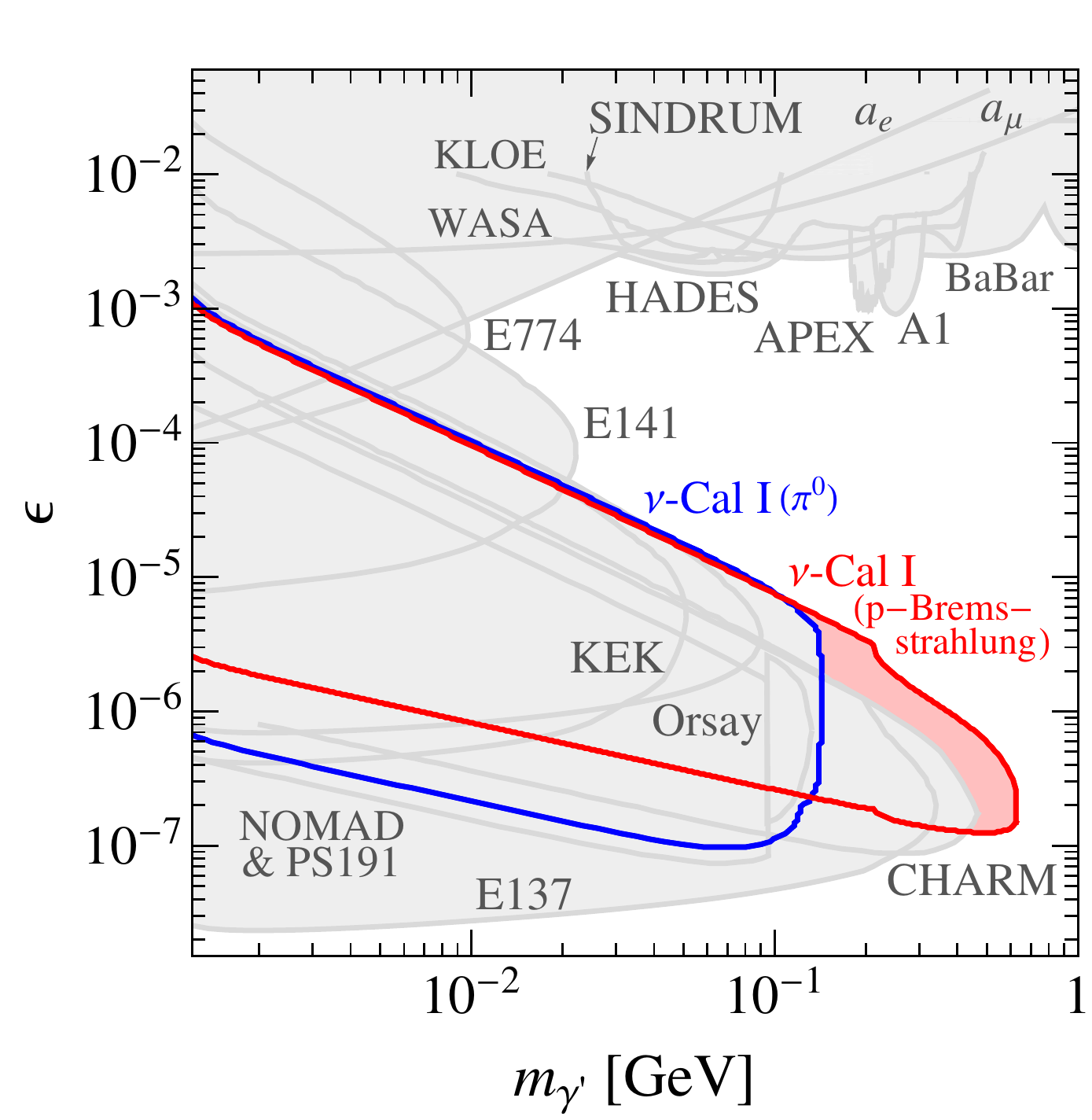}
\end{center}
\caption[]{
\small
Comparison of the present exclusion bounds (red area and lines) with other exclusion limits (grey area and 
lines) derived from data on the anomalous magnetic moments of the electron and muon \cite{Pospelov:2008xx},
$\Upsilon(3S)$ decays \cite{UPS}, 
Belle \cite{BELLE}, 
$J/\psi$-decays \cite{JPSI}, 
$K$-decays \cite{Beranek:2012ey},  
KLOE-2 \cite{Babusci:2012cr},
A1 \cite{Merkel:2011ze}, 
APEX \cite{Abrahamyan:2011gv}, 
HADES \cite{Agakishiev:2013jla},
E774 \cite{E774}, 
E141 \cite{E141}, 
E137 \cite{E137,Essig:2010xa}, 
Orsay \cite{Davier:1989wz}, 
KEK \cite{Konaka:1986cb}, 
NOMAD and PS191 \cite{Gninenko:2011uv}, 
CHARM \cite{Gninenko:2012eq}, 
SINDRUM \cite{Gninenko:2013sr},
WASA \cite{Adlarson:2013eza} and
$\nu$-CAL I for $\pi_0$-decay \cite{Blumlein:2011mv} (blue line).${}^\thefootnote$
} 
\label{FIG:5} 
\end{figure}
\end{center}

Using the pair production channel, the present analysis is sensitive to $\gamma'$ particles in the approximate range of
$5\cdot 10^{-5} < \epsilon < 10^{-1}$ and 0.1~eV$ < m_{\gamma'} < $1~MeV. Confronting this 
with a summary exclusion plot as 
Figure~4 of~\cite{Jaeckel:2010xx}, this range has already been excluded by different methods such as solar 
lifetime considerations and 
precision measurements of the Rydberg constant. However, the value of the present analysis consists in using a direct detection 
method which is largely model independent. 

Confidence limits are calculated with the CLs method~\cite{CL} according to
\begin{equation}
c=1-\sum_{n=0}^{N}P(n,s+b)/\sum_{n=0}^{N}P(n,b)~,
\label{Eq:CL}
\end{equation}
with $P(n,x)$ the Poisson-probability to observe $n$ events for a mean value of $x$. $N$ denotes the 
number of events being actually observed
and $b$ is the background estimate from simulations. Based on these values Eq.~(\ref{Eq:CL}) 
allows to determine the signal level $s$ for a given confidence level.
For $N=1$ observed events and a background of $b=0.3$ ($e^+e^-$ channel)
a signal of 4.5 events can be excluded at 95\% C.L. $(c = 0.95)$. 
This value changes to 4.7~events if we conservatively assume an uncertainty  
of a factor two for the background estimate $b$.
For the muon channel with $N=1$ and $b=0.7$ we obtain a 95\% C.L. limit
of 4.5 events when assuming the same uncertainty for the background estimate.

If $k$ different channels are combined such as the decays into electrons and muons Eq.~(\ref{Eq:CL}) 
modifies to
\begin{equation}
c=1-\prod_{k=1}^K\left[\sum_{n=0}^{N}P(n,s_k+b_k)\right]/\prod_{k=1}^K\left[\sum_{n=0}^{N}P(n,b_k)\right]~.
\end{equation}
This relation is used to calculate the corresponding event numbers for both muon and electron signatures 
at the 95\% C.L. for the mass range where both channels contribute.

The new corresponding exclusion region is shown as red area (and line) 
in comparison with the limit from~\cite{Blumlein:2011mv} (blue line)
and limits from other experiments in Figure~\ref{FIG:5}. 

At large values of $\epsilon$ studies of the anomalous magnetic moments of the muon and 
electron~\cite{Pospelov:2008xx},
of rare decays of heavy mesons~\cite{UPS}, and results from MAMI \cite{Merkel:2011ze},  
put stringent limits. For 
$10^{-3} < \epsilon < 10^{-7}$ beam dump 
experiments~\cite{E137,E141,E774,Davier:1989wz,Konaka:1986cb,Gninenko:2011uv,Gninenko:2012eq,Blumlein:2011mv} 
give the
best sensitivity. For even smaller values of $\epsilon$ limits can be derived
by studying the dynamics of supernovae cooling~\cite{Turner:1987by}. 
The prospects for the sensitivity of a reanalysis of LSND 
data~\cite{LSND} have been noted in \cite{Essig:2010gu} earlier, cf. also
the summary in \cite{Blumlein:2011mv}, Figure~5.

\section{Conclusions}

\vspace*{1mm}
\noindent
We have re-analyzed proton beam dump data taken at the U70 accelerator at IHEP
Serpukhov with the $\nu$-calorimeter I experiment in 1989 
\cite{Blumlein:1990ay,Blumlein:1991xh} to set mass and coupling limits for
dark gauge forces. The search is based on $\gamma'$-Bremsstrahlung off the 
incoming proton beam searching for electromagnetic showers and muon pairs 
in a neutrino calorimeter \cite{Barabash:2002zd}. 
Recently published limits based on the same dataset~\cite{Blumlein:2011mv} could 
be extended  towards larger gauge boson masses, exclucing a new area in the 
$m_{\gamma'}-\epsilon$ plane.
The present analysis extends the region excluded by a recently published limit 
based on the same dataset~\cite{Blumlein:2011mv} 
towards larger masses $m_{\gamma'} \in [m_{\pi_0} , 0.63~{\rm GeV} ]$ for
values in the mixing parameter $\epsilon \approx 10^{-6}$. 
In future experiments signals from dark gauge forces 
will be searched for in the yet unexplored regions shown in Figure~\ref{FIG:5}, see e.g. Ref.~\cite{Andreas:2010tp} 
for proposals.

\vspace*{4mm}
\noindent
{\bf Acknowledgment.}~
{We would like to thank our former colleagues of the $\nu$-Cal I experiment 
for collaboration during the time 1983--1990. We thank S.~Andreas, D.Yu.~Bardin, 
P.V.~Landshoff, T.~Riemann, A. Ringwald, M.~Walter and M.~Veltman for discussions.
This paper has been supported in part by DFG Sonderforschungsbereich Transregio 9, 
Computergest\"utzte Theoretische Teilchenphysik and EU Network {\sf LHCPHENOnet} 
PITN-GA-2010-264564.}


\begin{thebibliography}{99}
\footnotesize
%
\bibitem{Blumlein:1990ay}
  J.~Bl\"umlein, J.~Brunner, H.~J.~Grabosch, P.~Lanius, S.~Nowak, C.~Rethfeldt,
  H.~E.~Ryseck, M.~Walter {\it et al.},
  Z.\ Phys.\  {\bf C51 } (1991)  341.
%
\bibitem{Blumlein:1991xh}
  J.~Bl\"umlein, J.~Brunner, H.~J.~Grabosch, P.~Lanius, S.~Nowak, C.~Rethfeldt,
  H.~E.~Ryseck, M.~Walter {\it et al.},
  Int.\ J.\ Mod.\ Phys.\  {\bf A7 } (1992)  3835.
%
\bibitem{Blumlein:2011mv}
  J.~Bl\"umlein and J.~Brunner,
  Phys.\ Lett.\ B {\bf 701} (2011) 155
  [arXiv:1104.2747 [hep-ex]].
%
\bibitem{FAYET}
  P.~Fayet,
  Phys.\ Lett.\  {\bf B96 } (1980)  83;
  Nucl.\ Phys.\  {\bf B187 } (1981)  184.
%
\bibitem{Holdom:1985ag}
  B.~Holdom,
  Phys.\ Lett.\  {\bf B166 } (1986)  196.
%
\bibitem{Fayet:1990wx}
  P.~Fayet,
  Nucl.\ Phys.\  {\bf B347 } (1990)  743; 
  Phys.\ Lett.\  {\bf B675 } (2009)  267. 
  [arXiv:0812.3980 [hep-ph]].
%
\bibitem{Redondo:2010dp}
  J.~Redondo and A.~Ringwald,
  Contemp.\ Phys.\  {\bf 52} (2011) 211
  [arXiv:1011.3741 [hep-ph]].
%
\bibitem{Andreas:2010tp}
  S.~Andreas, A.~Ringwald,
  [arXiv:1008.4519 [hep-ph]].
%
\bibitem{FELDMAN}
  D.~Feldman, B.~Kors, P.~Nath,
  Phys.\ Rev.\  {\bf D75 } (2007)  023503,
  [hep-ph/0610133];\\
  D.~Feldman, Z.~Liu, P.~Nath,
  Phys.\ Rev.\  {\bf D75 } (2007)  115001,
  [hep-ph/0702123].
%
\bibitem{Essig:2010gu}
  R.~Essig, R.~Harnik, J.~Kaplan, N.~Toro,
  Phys.\ Rev.\  {\bf D82 } (2010)  113008,
  [arXiv:1008.0636 [hep-ph]].
%
\bibitem{MILLI}
  S.~Davidson, S.~Hannestad, G.~Raffelt,
  JHEP {\bf 0005 } (2000)  003,
  [hep-ph/0001179];\\
  M.~Gl\"uck, S.~Rakshit, E.~Reya,
  Phys.\ Rev.\  {\bf D76 } (2007)  091701,
  [hep-ph/0703140];\\ 
  S.~L.~Adler, J.~Gamboa, F.~Mendez, J.~Lopez-Sarrion,
  Annals Phys.\  {\bf 323 } (2008)  2851, 
  [arXiv:0801.4739 [hep-ph]].
%
\bibitem{Bjorken:2009mm}
  J.~D.~Bjorken, R.~Essig, P.~Schuster, N.~Toro,
  Phys.\ Rev.\  {\bf D80 } (2009)  075018,
  [arXiv:0906.0580 [hep-ph]].
%
\bibitem{Stueckelberg:1938}
  E.~C.~G.~Stueckelberg,
  Helv.\ Phys.\ Acta {\bf 11 } (1938)  225;
~299;
~312.
%
\bibitem{Pospelov:2008xx}
  M.~Pospelov,
  Phys.\ Rev.\  {\bf D80 } (2009)  095002,
  [arXiv:0811.1030 [hep-ph]];\\
  H.~Davoudiasl, H.-S.~Lee and W.~J.~Marciano,
  Phys.\ Rev.\ D {\bf 86} (2012) 095009
  [arXiv:1208.2973 [hep-ph]];\\
  M.~Endo, K.~Hamaguchi and G.~Mishima,
  Phys.\ Rev.\ D {\bf 86} (2012) 095029
  [arXiv:1209.2558 [hep-ph]].
%
\bibitem{UPS}
  R.~Essig, P.~Schuster, N.~Toro,
  Phys.\ Rev.\  {\bf D80 } (2009)  015003.
  [arXiv:0903.3941 [hep-ph]];\\
  B.~Aubert {\it et al.} [ BABAR Collaboration ],
  [arXiv:0902.2176 [hep-ex]].
%
\bibitem{BELLE}
I.~Jaegle, Belle-collab., Frascati Phys. Ser. {\bf LVI} (2012) 131.
%
\bibitem{JPSI}
J.~Fu, BES-III-collab., Frascati Phys. Ser. {\bf LVI} (2012) 122.
%
\bibitem{Beranek:2012ey}
  T.~Beranek and M.~Vanderhaeghen,
  Phys.\ Rev.\ D {\bf 87} (2013) 015024
  [arXiv:1209.4561 [hep-ph]].
%
\bibitem{Babusci:2012cr}
  D.~Babusci {\it et al.}  [KLOE-2 Collaboration],
  Phys.\ Lett.\ B {\bf 720} (2013) 111
  [arXiv:1210.3927 [hep-ex]];\\
  L.~Barze, G.~Balossini, C.~Bignamini, C.~M.~C.~Calame, G.~Montagna, O.~Nicrosini and F.~Piccinini,
  Eur.\ Phys.\ J.\ C {\bf 71} (2011) 1680
  [arXiv:1007.4984 [hep-ph]];
  Acta Phys.\ Polon.\ B {\bf 42} (2011) 2461.
%
\bibitem{Merkel:2011ze}
  H.~Merkel {\it et al.}  [A1 Collaboration],
  Phys.\ Rev.\ Lett.\  {\bf 106} (2011) 251802
  [arXiv:1101.4091 [nucl-ex]].
%
\bibitem{Abrahamyan:2011gv}
  S.~Abrahamyan {\it et al.}  [APEX Collaboration],
  Phys.\ Rev.\ Lett.\  {\bf 107} (2011) 191804
  [arXiv:1108.2750 [hep-ex]];\\
  {\tt http://hallaweb.jlab.org/experiment/APEX/}.
%
\bibitem{Agakishiev:2013jla}
  G.~Agakishiev {\it et al.}  [ HADES Collaboration],
  arXiv:1311.0216 [hep-ex].
%
\bibitem{E774}
  A.~Bross, M.~Crisler, S.~H.~Pordes, J.~Volk, S.~Errede, J.~Wrbanek,
  E774 collaboration,  
  Phys.\ Rev.\ Lett.\  {\bf 67 } (1991)  2942. 
%
\bibitem{E141}
  E.~M.~Riordan, M.~W.~Krasny, K.~Lang, P.~De Barbaro, A.~Bodek, S.~Dasu, N.~Varelas, X.~Wang {\it et al.},
  E141 collaboration,  
  Phys.\ Rev.\ Lett.\  {\bf 59 } (1987)  755.
%
\bibitem{E137}
  J.~D.~Bjorken, S.~Ecklund, W.~R.~Nelson, A.~Abashian, C.~Church, B.~Lu, L.~W.~Mo, T.~A.~Nunamaker {\it et al.},
  E137 collaboration,  
  Phys.\ Rev.\  {\bf D38 } (1988)  3375.
%
\bibitem{Essig:2010xa}
  R.~Essig, P.~Schuster, N.~Toro and B.~Wojtsekhowski,
  JHEP {\bf 1102} (2011) 009
  [arXiv:1001.2557 [hep-ph]].
%
\bibitem{Davier:1989wz}
  M.~Davier and H.~Nguyen Ngoc,
  Phys.\ Lett.\ B {\bf 229} (1989) 150.
%
\bibitem{Konaka:1986cb}
A.~Konaka, K.~Imai, H.~Kobayashi, A.~Masaike, K.~Miyake, T.~Nakamura, N.~Nagamine and N.~Sasao {\it et al.},
  Phys.\ Rev.\ Lett.\  {\bf 57} (1986) 659.
%
\bibitem{Gninenko:2011uv}
  S.~N.~Gninenko,
  Phys.\ Rev.\ D {\bf 85} (2012) 055027
  [arXiv:1112.5438 [hep-ph]].
%
\bibitem{Gninenko:2012eq}
  S.~N.~Gninenko,
  Phys.\ Lett.\ B {\bf 713} (2012) 244
  [arXiv:1204.3583 [hep-ph]].
%
\bibitem{Gninenko:2013sr}
  S.~N.~Gninenko,
  Phys.\ Rev.\ D {\bf 87} (2013) 035030
  [arXiv:1301.7555 [hep-ph]].
%
\bibitem{Adlarson:2013eza}
  P.~Adlarson {\it et al.}  [WASA-at-COSY Collaboration],
  Phys.\  Lett.\  B {\bf 726} (2013) , 187
  [arXiv:1304.0671 [hep-ex]].
%
\bibitem{Turner:1987by}
  M.~S.~Turner,
  Phys.\ Rev.\ Lett.\  {\bf 60 } (1988)  1797.
%
\bibitem{LSND}
  C.~Athanassopoulos {\it et al.} [ LSND Collaboration ],
  Phys.\ Rev.\  {\bf C58 } (1998)  2489, 
  [nucl-ex/9706006];\\
  L.~B.~Auerbach {\it et al.} [ LSND Collaboration ],
  Phys.\ Rev.\ Lett.\  {\bf 92 } (2004)  091801,
  [hep-ex/0310060].
%
\bibitem{Batell:2009di}
  B.~Batell, M.~Pospelov, A.~Ritz,
  Phys.\ Rev.\  {\bf D80 } (2009)  095024,
  [arXiv:0906.5614 [hep-ph]];\\
  H.~An, M.~Pospelov and J.~Pradler,
  Phys.\ Lett.\ B {\bf 725} (2013) 190
  [arXiv:1302.3884 [hep-ph]].
%
\bibitem{Dent:2012mx}
  J.~B.~Dent, F.~Ferrer and L.~M.~Krauss,
  arXiv:1201.2683 [astro-ph.CO].
%
\bibitem{Freytsis:2009bh}
  M.~Freytsis, G.~Ovanesyan and J.~Thaler,
  JHEP {\bf 1001} (2010) 111
  [arXiv:0909.2862 [hep-ph]];\\
J. Jaros, Frascati Phys. Ser. {\bf LVI} (2012) 33;\\
J.R. Boyce, Frascati Phys. Ser. {\bf LVI} (2012) 48;\\
  J.~Balewski, J.~Bernauer, W.~Bertozzi, J.~Bessuille, B.~Buck, R.~Cowan, K.~Dow and C.~Epstein {\it et al.},
  arXiv:1307.4432 [physics.ins-det].
%
\bibitem{KATZIN}
D.R. Katzin, {\sf The DarkLight Experiment: Searching for the Dark Photon}, Bachelor Thesis, MIT, June 2012.
%
\bibitem{Ebr:2013iea}
  J.~Ebr and P.~Ne\u{c}esal,
  Phys.\ Lett.\ B {\bf 725} (2013) 185
  [arXiv:1307.3890 [astro-ph.HE]].
%
\bibitem{Andreas:2012xh}
  S.~Andreas,
  arXiv:1211.5160 [hep-ph]; PhD Thesis \newline 
  {\tt http://www-library.desy.de/cgi-bin/showprep.pl?desy-thesis-13-024}.
%
\bibitem{Andreas:2012mt}
  S.~Andreas, C.~Niebuhr and A.~Ringwald,
  Phys.\ Rev.\ D {\bf 86} (2012) 095019
  [arXiv:1209.6083 [hep-ph]].
%
\bibitem{Reece:2009un}
  M.~Reece and L.-T.~Wang,
  JHEP {\bf 0907} (2009) 051
  [arXiv:0904.1743 [hep-ph]].
%
\bibitem{FRASCATI}
Proceedings of DARK2012 {\sf Dark Forces at Accelerators}, Oct 16-19, 2012.~Frascati Phys. Ser. {\bf LVI} 
Eds.~F.~Bossi, S. Giovannella, P. Santangelo, B. Sciascia, ISBN 978-88-86409-62-9.
%
\bibitem{Gninenko:2013rka}
  S.~N.~Gninenko,
  arXiv:1308.6521 [hep-ph].
%
\bibitem{Gershtein:2013iqa}
  Y.~Gershtein, M.~Luty, M.~Narain, L.~-T.~Wang, D.~Whiteson, K.~Agashe, L.~Apanasevich and G.~Artoni {\it et al.},
  arXiv:1311.0299 [hep-ex].
%
\bibitem{Barabash:2002zd}
  L.~S.~Barabash, S.~A.~Baranov, Y.~A.~Batusov, S.~A.~Bunyatov, V.~Y.~.Valuev, I.~A.~Golutvin, 
  O.~Y.~.Denisov and M.~Y.~.Kazarinov {\it et al.},
  Instrum.\ Exp.\ Tech.\  {\bf 46} (2003) 300
   [Prib.\ Tekh.\ Eksp.\  {\bf 46} (2003) 20].
%
\bibitem{AXION}
  P.~Fayet,
  {\sf Introduction To Axions}, in:
  Proceedings of the Conference 
``QCD and Lepton Physics'', Vol. 1, pp.~307-313, (Ecole Norm. Sup., Paris, 1981), LPTENS 81-08;
\\
  R.~D.~Peccei,
  Adv.\ Ser.\ Direct.\ High Energy Phys.\  {\bf 3 } (1989)  503; 
  Lect.\ Notes Phys.\  {\bf 741 } (2008)  3, 
  [hep-ph/0607268];\\
  J.~Bl\"umlein, {\sf On the experimental status of light pseudo-scalar particles},
{ Proceedings of the High Energy Physics Workshop Georgenthal, Germany 
(Leipzig, University Press, 1984), p.~100.
}
\\
  L.~Barabash, S.~Baranov, Y.~A.Batusov, S.~Bunyatov, O.~Denisov, A.~Karev, M.~Kazarinov, O.~Klimov {\it et 
  al.},
  Phys.\ Lett.\  {\bf B295 } (1992)  154; Sov.\ J.\ Nucl.\ Phys.\  {\bf 55 } (1992)  
  1810;\\
  P.~Sikivie,
  Nucl.\ Phys.\ Proc.\ Suppl.\  {\bf 87 } (2000)  41, 
  [hep-ph/0002154];\\
  J.~E.~Kim, G.~Carosi,
  Rev.\ Mod.\ Phys.\  {\bf 82 } (2010)  557, 
  [arXiv:0807.3125 [hep-ph]].
%
\bibitem{PROP}
J. Bl\"umlein, R. Nahnhauer, S. Nowak, S. Schlenstedt, and M. Walter, {
\sf Estimation of the Axion Determination Rate via the Decay $a^0 \rightarrow 
\gamma \gamma$ in a Proton Beam-Dump Experiment at the U-70-Accelerator}, internal 
note, IfH AdW, January 1984,~31~p. (unpublished).
%
\bibitem{Fermi:1924tc}
  E.~Fermi,
  {Z.\ Phys.}\  {\bf 29} (1924) 315.
%
\bibitem{Williams}
E.J. Williams, {Proc. Roy. Soc. London} (A) {\bf 139} (1933) 163;
  {Phys.\ Rev.}\  {\bf 45} (1934) 729;
Mat. Fys. Medd. {\bf 13} (1935) 4.
%
\bibitem{vonWeizsacker:1934sx}
  C.~F.~von Weizs\"acker, 
  {Z.\ Phys.}\  {\bf 88} (1934) 612.
%
\bibitem{Kessler:1975}
P. Kessler, 
Acta Phys. Austr. {\bf 41} (1975) 141 and references therein.
%
\bibitem{Kessler:1960}
P. Kessler, { Nuovo Cim.} {\bf 17} (1960) 809.
%
\bibitem{LU}
K. Logar and P. Urban,
Sitzungsber. der \"Osterreichischen Akademie der Wissenschaften, {\bf 171} 
(1962) 197;
~Ann. Phys. (Leipzig) {\bf 11} (1963) 101.
%
\bibitem{Weinberg:1966jm}
  S.~Weinberg,
  Phys.\ Rev.\  {\bf 150} (1966) 1313; Erratum:  {\bf 158} (1967) 1638.\\
  A. Ramakrishnan, J. K. Radha, and R. Thunga, Proc. Indian Acad. Sci.
  {\bf 52}(A) (1960) 228; J. Math. Anal. Appl. {\bf 4}
  (1962) 494; {\bf 5} (1962) 225; \\ 
  A. Ramkrishnan, K. Venkatesan, and V. Devanathan, J. Math. Anal. Appl. {\bf 8} (1964) 
  345.
%
\bibitem{Baier:1973ms}
  V.~N.~Baier, V.~S.~Fadin and V.~A.~Khoze,
  Nucl.\ Phys.\ B {\bf 65} (1973) 381.  
%
\bibitem{Chen:1975sh}
  M.-S.~Chen and P.~M.~Zerwas,
  Phys.\ Rev.\ D {\bf 12} (1975) 187.
%
\bibitem{Altarelli:1977zs}
  G.~Altarelli and G.~Parisi,
  Nucl.\ Phys.\ B {\bf 126} (1977) 298.
%
\bibitem{Martin:1996eva}
  A.~D.~Martin, R.~G.~Roberts, M.~G.~Ryskin and W.~J.~Stirling,
  Eur.\ Phys.\ J.\ C {\bf 2} (1998) 287
  [hep-ph/9612449].
%
\bibitem{Kim:1973he}
  K.~J.~Kim and Y.~-S.~Tsai,
  Phys.\ Rev.\ D {\bf 8} (1973) 3109.
%
\bibitem{Frixione:1993yw}
  S.~Frixione, M.~L.~Mangano, P.~Nason and G.~Ridolfi,
  Phys.\ Lett.\ B {\bf 319} (1993) 339
  [hep-ph/9310350].
%
\bibitem{Vermaseren:2000nd}
  J.~A.~M.~Vermaseren,
  math-ph/0010025.
%
\bibitem{Proca:1936}
A.~Proca, J. de Phys. et le Radium {\bf 7} (1936) 347.
%
\bibitem{CJ:1950}
F. Coester and J.M. Jauch 
Phys. Rev. {\bf D78} (1950) 149.
%
\bibitem{Coester:1951}
F. Coester, 
Phys. Rev. {\bf D83} (1951) 798.
%
\bibitem{JR:1976}
J.M.~Jauch and F.~Rohrlich, 
{\sf The Theory of Photons and Electrons: The Relativistic Quantum Field Theory 
of Charged Particles with Spin One-half}, (Springer, Berlin, 1976).
%
\bibitem{Ruegg:2003ps}
  H.~Ruegg and M.~Ruiz-Altaba,
  Int.\ J.\ Mod.\ Phys.\ A {\bf 19} (2004) 3265
  [hep-th/0304245].
%
\bibitem{Linsker:1972dn}
  R.~Linsker,
  Phys.\ Rev.\ D {\bf 5} (1972) 1709.
%
\bibitem{Akhundov:1976yy}
  A.~A.~Akhundov and D.~Yu.~Bardin,
  JINR-P2-9587.
%
\bibitem{Nakamura:2010zzi}
K. Nakamura et al. (Particle Data Group), J. Phys. {\bf G 37} (2010) 075021; \newline
{\tt http://pdg.lbl.gov/}
%
\bibitem{Kim:1972gw}
  K.~J.~Kim and Y.~-S.~Tsai,
  Phys.\ Lett.\ B {\bf 40} (1972) 665.
%
\bibitem{Jaeckel:2010xx}
 J. J\"ackel and A. Ringwald,
 arXiv:1002.0329v1 [hep-ph].
%
\bibitem{CL} G. Zech, Nucl.\ Instr.\ Meth. {\bf A277} (1989) 608.
\end{thebibliography}
\end{document}